# Collective octahedral tilting in ultrathin Ruddlesden-Popper perovskite under terahertz light

Kun Liu, Jian Zhou[*]

*Center for Alloy Innovation and Design, State Key Laboratory for Mechanical Behavior of Materials, Xi'an Jiaotong University, Xi'an 710049, China*

Abstract

Perovskites have been applied in a wide range of fields such as solar cells and non-volatile memories due to their multiferroic nature and excellent photo-electric conversion capabilities. Recently, two-dimensional (2D) perovskites with a few atomic layers have been successfully synthesized, attracting significant attention for potential applications. In this work, we perform first-principles calculations to investigate an ultrathin prototypical Ruddlesden−Popper phase, $Bi_2FeO_4$, with its thickness down to one unit cell. We show that this compound could exist in two (meta-)stable octahedral tilting phases, belonging to *P*$2_1$/*c* and *C*2/*m* space groups, respectively. Using the optomechanical theory, we suggest that reversible and non-volatile phase switching can be triggered using non-destructive terahertz light. In addition, the two phases show distinct optical reflectance spectrum in the visible light range, which can be used as an optical probe for phase transformation. This enables both "write" and "read" in an all-optical route.

[*] Email: jianzhou@xjtu.edu.cn

Perovskites are promising materials that can be used for energy storage and conversion,[1,2] sensor devices,[3,4] and non-volatile memories,[5,6] etc. While three-dimensional bulk perovskites ($ABX_3$ with X being a halogen or oxygen element) have been extensively studied and widely used, 2D perovskite thin films have also been attracting tremendous attention due to their enhanced information storage density,[7] flexible mechanical deformation ability,[8] and distinct electric/optical feature.[9] Owing to the high surface-to-volume ratio, 2D perovskites are more amenable to external stimuli, such as mechanical strain and light exposure.[10-12] For example, Ji *et al.* have recently synthesized freestanding $BiFeO_3$ and $SrTiO_3$ thin films with thickness down to a few atomic layers, sparking new insights into ultrathin perovskites.[13]

In this Letter, we design and explore the phase stability and transformation of a 2D perovskite with 1 unit cell (U.C.) thickness, which is the thinnest limit in the Ruddlesden-Popper (RP) phase.[9,14] Currently, such single-layer $Bi_2FeO_4$ could be fabricated using pulsed laser deposition or molecular beam epitaxy methods. These methods have been adopted to synthesize high-quality RP phase perovskite.[15-19] According to first-principles density functional theory (DFT) calculations, we show that this material could exist in two different phases with distinct $FeO_6$ octahedral tilts. They are separated and protected by an obvious energy barrier (calculated to be ~142 meV/f.u., or 14.7 $\mu J/cm^2$), and can transform between each other under terahertz light irradiation. The contrasting electronic structures also ensure that these two phases can be detected and distinguished by their reflectance spectrum at the visible light regime. Hence, we suggest that the 1 U.C. thick RP perovskite could serve as non-volatile phase transformation materials with the information "read" and "write" processes through an all-optical scheme. The noncontacting nature of optical manipulation and detection could greatly reduce the lattice damage and potential foreign atom introduction into the systems.

Our DFT calculations are performed in the Vienna *ab initio* simulation package (VASP), in which the exchange-correlation potential is treated by the generalized gradient approximation in the Perdew–Burke–Ernzerhof form.[20,21] We use the projector augmented-wave method[21,22] to describe the core electrons and the planewave basis set to expand the valence electrons, with a kinetic cutoff energy of 500 eV. The strong correlation effect on the Fe-$d$ orbitals is treated by introducing an on-site Hubbard $U$ energy[23,24] with an effective value of 4 eV, in accordance with previous works.[25,26] Note that we have tested other $U$ values, which yield qualitatively the same results. In order to eliminate the periodic image interactions, we add a vacuum space of over 12 Å in the out-of-plane $z$ direction. The Monkhorst−Pack scheme[27] is used to represent the first Brillouin zone (BZ) with the $k$-mesh grids denser than 0.04 Å$^{-1}$ along each Cartesian axis. During the geometric optimization, both the atomic internal coordinates and the lattice parameters have been well relaxed. The convergence threshold for the total energy and force components are set to be $10^{-6}$ eV and $10^{-2}$ eV/Å, respectively. Phonon dispersion is calculated via the finite displacement method.[28] The spin-orbit coupling (SOC) is self-consistently included in the electronic structure calculations. We also adopt the hybrid functional (in the form of Heyd−Scuseria−Ernzerhof, HSE06) treatment[29] for better electronic dispersion and optical response function predictions.

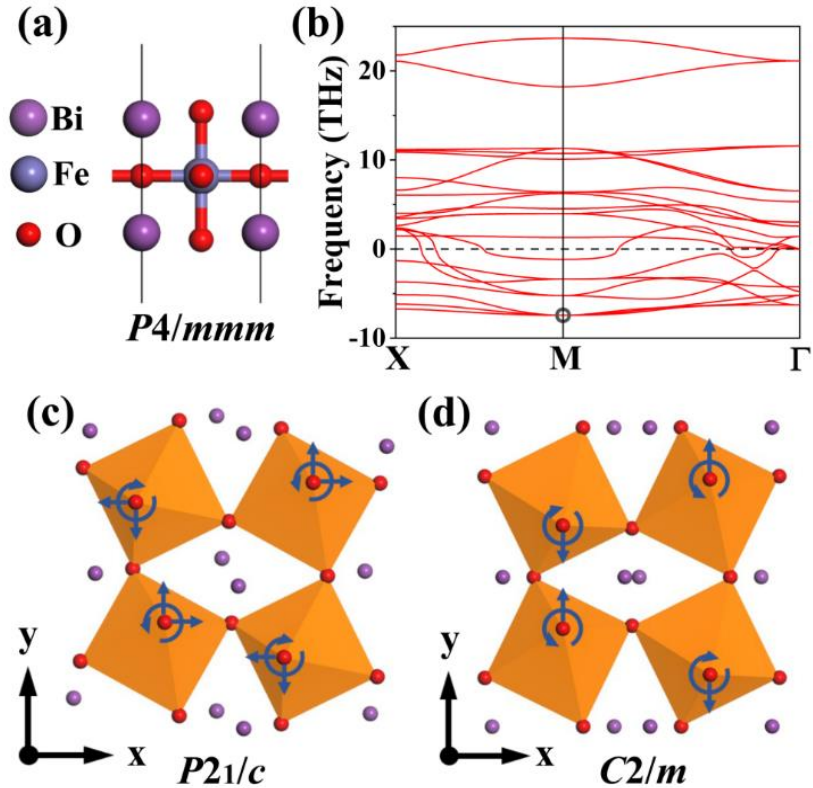

Figure 1. (a) The atomic structure and (b) phonon dispersion of high symmetric $P4/mmm$ phase. Geometric structure of lower symmetric (c) $P2_1/c$ and (d) $C2/m$, in which we show the octahedral rotation and tilt using circular and straight arrows,

respectively.

The chemical formula of a typical 2D RP perovskite is $A_{n+1}B_nO_{3n+1}$, where $n$ represents the unit cell number. Obviously, the thinnest case corresponds to $n = 1$, giving $A_2^{3+}B^{2+}O_4^{2-}$. We plot the geometric structure of $Bi_2FeO_4$ in Fig. 1(a), where the atoms reside at their high symmetric positions, belonging to a layer group of *P*4/*mmm*. We perform fixed symmetry geometric structure optimization for the high symmetric structure, yielding its lattice constant of 3.82 Å. In order to explore its dynamic stability, we calculate its phonon dispersion, and observe several imaginary branches [Fig. 1(b)]. Then we perform careful structural relaxation according to these displacement modes. In this process, the soft modes at Γ finally become to either the high symmetry phase (*P*4/*mmm*) or structural destruction, and do not yield a proper ferroelectric phase transition. Note that in 2D perovskites, there are two types of octahedral rotational modes, namely, a rotation along the out-of-plane axis and an in-plane tilting. The three lowest imaginary modes at the *M* point (corner of the BZ) correspond to one $FeO_6$ octahedral rotation mode and two tilting modes (Fig. S1 of Supplemental Material, SM), respectively.

We follow these octahedral rotation and tilting modes to build several low symmetric phases (Table 1). After geometric relaxation with fixed symmetry, we finally yield two phases, namely, $a^-a^-c$ and $a^-b^0c$ in the Glazer notation.[30] The *a*, *b* and *c* represent the axes of octahedral rotation/tilting along the [100], [010] and [001] directions, respectively. The subscripts "−" and "0" denote anti-phase tilting and no tilting along the axes, respectively. After careful relaxation, they lead to either unstable structures or one of these two phases. Compared with the *P*4/*mmm*, both of them possess octahedral rotations, while they exhibit biaxial (*P*2$_1$/*c*) and uniaxial (*C*2/*m*) octahedral tilting. Their optimized geometries are shown in Figs. 1(c) and 1(d). Since the number of A-O atomic layers is even, it is difficult to yield hybrid improper (anti-)ferroelectricity in a system with $n = 1$.[31] For the $Bi_2FeO_4$, both *P*2$_1$/*c* and *C*2/*m* are paraelectric phases. We find that the *P*2$_1$/*c* has a lower total energy than *C*2/*m* by 0.33 eV/f.u. (or 34.5 μJ/cm$^2$). In order to check their dynamic stability, we perform phonon calculations and find no imaginary modes (Fig. S2). In addition, we simulate *ab initio* molecular dynamics (AIMD) to check their thermal stability at 300 K. The structures do not show significant distortions after 10 ps duration (Fig. S3), suggesting that both $P2_1/c$ and $C2/m$ are stable under room temperature. Note that the bare surface may subject to attract surface ligand (such as oxygens from the atmospheric environment) which could strongly affect oxidation state of Fe.[32,33] Our test calculations suggest that the $Bi_2FeO_4$ may exhibit absorb ligands and termination on surface (see SM). Therefore, in order to eliminate such problems, one has to apply buffer layers to protect the sample during synthesis, which has been well-applied in the current experimental advances.

Table 1. Various octahedral rotation/tilting of $Bi_2FeO_4$ and their corresponding layer groups, relative energies, lattice parameters, and dynamic stability. Symbol "-" indicates that the structure breaks after relaxation.

| Octahedral rotation | Layer group | Relative energy (meV/f.u.) | Lattice parameters (Å) | Dynamic stability |
|---|---|---|---|---|
| $a^0a^0c^0$ | $P4/mmm$ | 2622 | 3.820 | No |
| $a^0a^0c$ | $P4/mbm$ | 1689 | 7.683 | No |
| $a^-a^-c^0$ | $Pmna$ | - | - | - |
| $a^-b^0c^0$ | $Cmma$ | - | - | - |
| $a^-b^-c^0$ | $P2/c$ | - | - | - |
| $a^-a^-c$ | $P2_1/c$ | 0 | 8.249 | Yes |
| $a^-b^0c$ | $C2/m$ | 325 | 8.113, 8.378 | Yes |

The Fe atoms are in their 2+ oxidation state, leaving four *d*-electrons unpaired. Our calculations also confirm that each Fe carries ~4 $\mu_B$ local magnetic moments. In order to explore their magnetic coupling character, we calculate three cases for each of these two phases, namely, ferromagnetic (FM), stripe-antiferromagnetic (sAFM) and checkerboard-antiferromagnetic (cAFM) configurations (Fig. S4). Their relative energies are tabulated in Table S1. We find that in both phases, the cAFM magnetic configuration possesses the lowest energy. The spin density distributions of them are depicted in Fig. S4, which shows that the spin polarization is primarily localized around Fe atoms. By applying an anisotropic Heisenberg model,[34] $H = -J\sum_{\langle i,j \rangle} \vec{S}_i \cdot \vec{S}_j - \lambda \sum_i (S_i^z)^2$ (summation $\langle i,j \rangle$ runs over nearest neighbor magnetic sites), we estimate the exchange parameter $J$ to be −1.10 and −2.03 meV for $P2_1/c$ and $C2/m$ phases, respectively. We also find that the magnetic easy direction of $P2_1/c$ lies in the $x$-$y$ plane, while that of the $C2/m$ is out-of-plane ($z$), suggesting a phase-dependent magnetocrystalline anisotropy character. The single-ion anisotropic parameters $\lambda$ are calculated to be −0.27 and 0.28 meV for $P2_1/c$ and $C2/m$, respectively. By performing Monte-Carlo simulations on a (32×32) spin lattice (Fig. S5), we observe that their magnetic susceptibility exhibits a cusp at 42 K and 67 K, respectively, giving the estimated Néel temperatures.

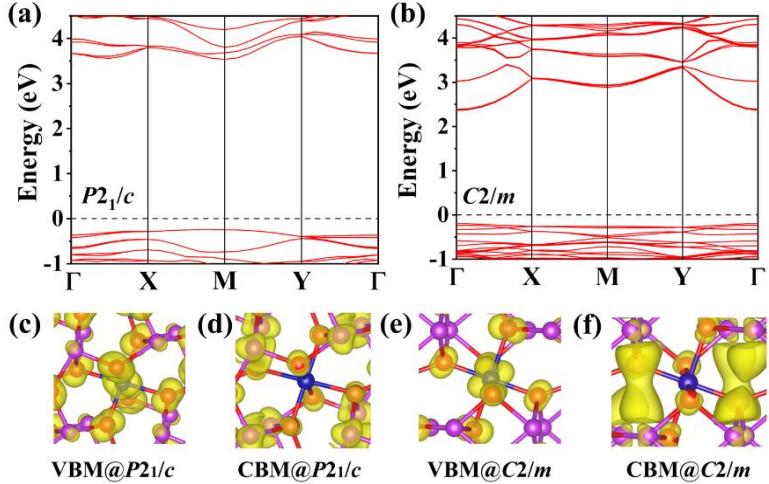

Figure 2. Hybrid functional (HSE06+SOC) calculated band dispersions for (a) $P2_1/c$ and (b) $C2/m$ phases. (c) – (f) Band-decomposed charge densities at the band edges.

Next, we adopt hybrid functional HSE06+SOC level of theory to evaluate their electronic band structures (at their ground cAFM configuration). As shown in Figs. 2(a) and 2(b), the calculated bandgaps are 3.8 ($P2_1/c$) and 2.6 ($C2/m$) eV, respectively. Both phases exhibit a direct bandgap character, with the valence band maximum (VBM) and conduction band minimum (CBM) locating at the $M$ ($P2_1/c$) and $\Gamma$ ($C2/m$) points. In order to explore their wavefunction components at the band edge, we compute their band-decomposed charge densities, as shown in Figs. 2(c)−2(f). The wavefunctions of the CBM distribution in the *C*2/*m* phase is delocalized (Fig. 3f), which is consistent with its band dispersion around Γ. The wavefunctions of the VBM in both phases are localized around the $FeO_6$ octahedrons, while those of the CBM are primarily distributed around the Bi and O atoms. These are consistent with the projected density of states results (Fig. S6). One notes that this band edge contribution character is different from the bulk $BiFeO_3$, in which the Fe-3*d* orbitals are located at the CBM.[35] This can be attributed to the valence state of $Fe^{2+}$ (rather than $Fe^{3+}$) in the RP phase, exhibiting a higher tendency to lose electrons.

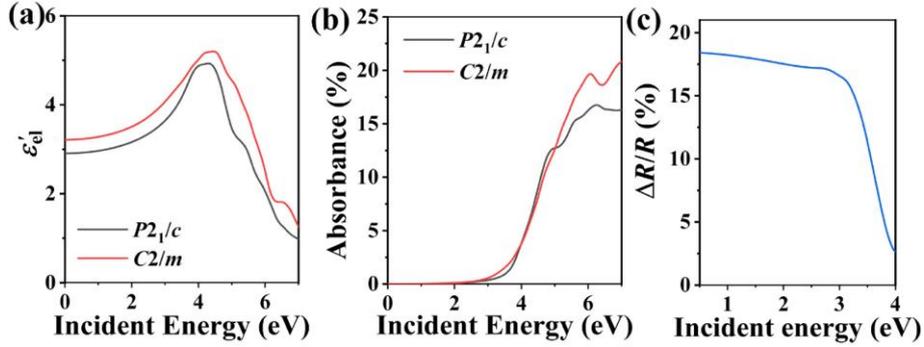

Figure 3. Electron contributed (ion-clamped) in-plane averaged (a) real part of dielectric functions and (b) the absorbance spectra for the *P*2$_1$/*c* and *C*2/*m* phases. (c) Relative difference in reflectance ($R$) between the two phases, $(R^{C2/m} - R^{P2_1/c})/R^{P2_1/c}$.

The contrasting electronic band structures between the two phases indicate distinct optical responses. We perform independent particle approximation (IPA) with the hybrid functional treatment to evaluate the electron contributed (ion-clamped) dielectric function[36]

$$\varepsilon_{ij}^{el}(\omega) = \delta_{ij} - \frac{e^2}{\varepsilon_0} \int_{BZ} \frac{d^3 k}{(2\pi)^3} \sum_{c,v} \frac{\langle u_{v,k}|\nabla_{k_i}|u_{c,k}\rangle \langle u_{c,k}|\nabla_{k_j}|u_{v,k}\rangle}{\hbar\left(\omega_{c,k}-\omega_{v,k}-\omega-\frac{i}{\tau^{el}}\right)} \quad (1)$$

where $|u_{n,k}\rangle$ is the periodic part of Bloch wavefunction for band *n* (*c* and *v* representing conduction and valence band, respectively) at ***k***. $\varepsilon_0$ is vacuum permittivity constant. $\tau^{el}$ denotes the electron relaxation lifetime, which is mainly determined by the quality of the sample, the electron-electron and electron-phonon scattering, etc. Even for perfect crystal, its evaluation is computationally challenging for each *n* and ***k***.

Here, we adopt the conventional approach to take a universal value of 0.03 ps. In general, its specific value does not strongly affect the main conclusion in this work, as we focus on the off-resonant frequency regime. The calculated in-plane averaged dielectric functions and absorbance spectra are plotted in Figs. 3(a) and 3(b), respectively. The absorbance spectra reveal that the maximum absorbance at the visible light frequency regime is about 20%. We then evaluate the reflectance spectrum using $R(\omega) = \frac{[N(\omega)-1]^2 + K(\omega)^2}{[N(\omega)+1]^2 + K(\omega)^2}$, in which $N(\omega)$ and $K(\omega)$ represent the frequency dependent refractive index and extinction coefficient, respectively. The results are shown in Fig. 3(c), showing an observable (~17%) reflectance contrast at the visible light frequency regime. This is sufficient to be distinguished in experiments.

In order to realize non-volatile information storage and memory, it is intriguing to explore the reversible phase transformation. By performing nudged elastic band calculation, we find that from $P2_1/c$ to $C2/m$ the energy barrier is 0.46 eV/f.u. (corresponding to FeO$_6$ octahedral tilting), and the energy barrier is 0.14 eV/f.u. from $C2/m$ to $P2_1/c$. The geometric structure of the saddle point phase is plotted in Fig. S7, which shows only one imaginary mode at Γ (Fig. S2). The energy barrier is moderate if one compares it with other well-studied phase change materials.[37-39] In practical applications, the energy barrier of phase change would be reduced under thermal effect and forming domain walls, etc.[37,39-41]

The thermodynamic phase transformation depends on the relative stability of Gibbs free energy (GFE). Here we discuss terahertz (THz) controlled phase transformation. Note that THz optics and related technology are intriguing owing to their unique advantages such as noncontacting and noninvasive nature, highly transparency character, and less susceptible to lattice damages.[42] Usually, the frequency regime emitted from THz sources is on the order of 1 THz,[43] which may interact with the material via both resonant and off-resonant approaches.[44] The resonant interaction corresponds to direct absorption of THz light, usually through exciting the infrared-active mode. It can be estimated via the imaginary parts of dielectric function, $A_{ii}(\omega) = 1 - e^{-\frac{\omega \varepsilon''_{ii}(\omega) d}{c_0}} \simeq \omega \varepsilon''_{ii}(\omega) d / c_0$,[45] where $\varepsilon''_{ii}(\omega)$ represents the imaginary part dielectric function of $i$-th diagonal component and $c_0$ is speed of light in vacuum. We use $d$ to denote the effective thickness of the 2D $Bi_2FeO_4$ layer (taken to be 5.5 Å), which also scales the dielectric function in the 2D limit. Hence, the choice of $d$ would not affect the GFE evaluation. The absorbance adds in the internal energy density by $ItA_{ii}(\omega)$, where $I$ is incident light intensity, $t$ refers to irradiation duration. The off-resonant interaction, on the other hand, refers to light scattering process that is related to the real part of dielectric function $\varepsilon'_{ii}(\omega)$. According to our previous work, it reduces the GFE density by $-\frac{d}{2c_0}I\varepsilon'_{ii}(\omega)$.[46] Hence, the total GFE density becomes (ignoring higher order interactions and thermal effect)[45]

$$\mathcal{G}(\omega) = \mathcal{G}_0 - \frac{d}{2c_0}I\varepsilon'_{ii}(\omega) + ItA(\omega) \qquad (2)$$

Here, we assume linearly polarized THz light (LPTL) irradiation, with its electric field component along the $i$-th direction. We consider a Gaussian form of the THz light

$$dI = I_0\rho(\omega)d\omega \qquad (3)$$

The probability density function $\rho(\omega) = \frac{1}{\sigma\sqrt{2\pi}}e^{-\frac{(\omega-\omega_0)^2}{2\sigma^2}}$, in which the $\sigma$ and $\omega_0$ are the frequency width at the half maximum and the light frequency center, respectively. Therefore, we obtain the total GFE density under THz illumination

$$\mathcal{G} = \mathcal{G}_0 + I_0\left[-\frac{d}{2c_0}\int \varepsilon'(\omega)\rho(\omega)d\omega + t\int A(\omega)\rho(\omega)d\omega\right] \qquad (4)$$

The first and second term inside the hard bracket represent the off-resonant scattering and resonant absorption effects, respectively. The detailed derivation can be found in SM. Here, we set $\sigma = 1$ THz and $t = 50$ ps, which is comparable with experimental setups.[47,48]

We check that the electronic bandgap remains to be large along the phase transformation reaction path (Fig. S8). Hence, under THz irradiation, there will be no electron-hole pair excitation during phase transformation, so that the electron contributed dielectric function $\varepsilon^{el}_{ij}(\omega)$ only affects the off-resonant scattering. We evaluate the ion contributed dielectric function according to the phonon spectrum and vibrational characters in the long wavelength limit ($\mathbf{q} \simeq 0$). According to linear response theory, it takes the form[49]

$$\varepsilon_{ij}^{\text{ion}}(\omega) = \frac{1}{V} \sum_m \frac{\mathcal{Z}_{m,i}^* \mathcal{Z}_{m,j}^*}{\omega_m^2 - \left(\omega + \frac{i}{\tau^{\text{ion}}}\right)^2} \quad (5)$$

where $\mathcal{Z}_{m,i}^* = \sum_{\kappa,i'} z_{\kappa,ii'}^* u_{m,\kappa i'}$ is the *i*-th Born effective charge of mode *m*, which quantifies the coupling between the infrared-active phonons and the electric fields. $\kappa$ is ion index and *u* is mass normalized displacement. $z_{\kappa,ii'}^*$ is the Born effective charge tensor component of ion-$\kappa$, which measures the change of polarization along the *i*-component and the ionic displacement in the *i'*-direction. The $\tau^{\text{ion}}$ is the phonon lifetime, and here we set $\tau^{\text{ion}} = 8$ ps, which is conservable in most cases.[50]

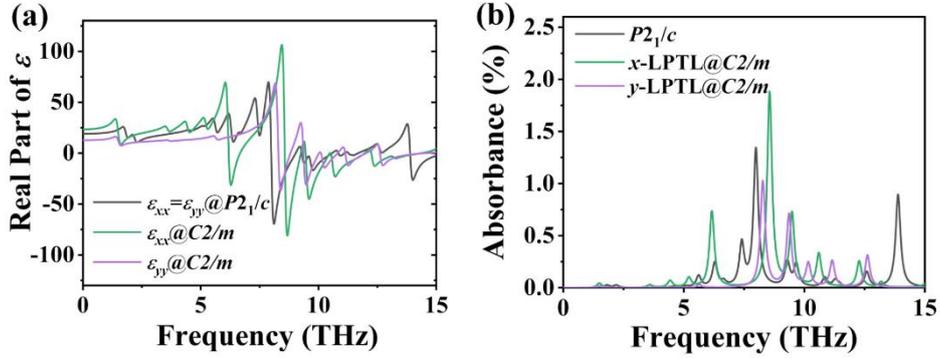

Figure 4. (a) Real part of dielectric function and (b) absorbance spectrum in the THz region. The *P*2$_1$/*c* phase response is nearly isotropic in the *x*-*y* plane.

By adding Eqs. (1) and (5), we calculate the total dielectric functions of *P*2$_1$/*c* and *C*2/*m* phases in THz region. The real part of dielectric functions is shown in Fig. 4(a). In the range of 5-10 THz, both two phases have a strong response to the LPTL. The absorbance spectra [Fig. 4(b)] reveals that the maximum absorbance in the THz region is ~2%. This marginal light absorption arises from the ultrathin nature of the material, so that the lattice damage and heat generation can be largely reduced.

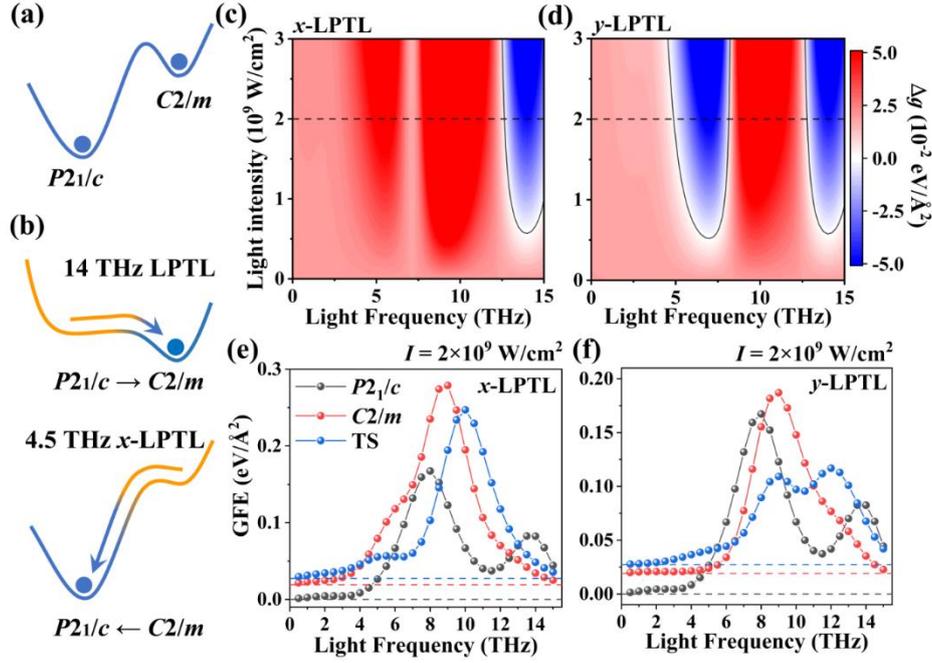

Figure 5. Schematic energy profile of $Bi_2FeO_4$ monolayer (a) without and (b) with THz illumination. (c) and (d) are the light-induced phase diagram under light frequency and intensity, measured by GFE density ($\Delta g = g_{C2/m} - g_{P2_1/c}$). (e) and (f) are the GFE density of two phases and transition state (denoted as TS) under $x$- and $y$-LPTL with the intensity of $2\times10^9$ W/cm$^2$ [the horizontal dotted lines in (c) and (d)], respectively.

Reversible phase transformation between the two phases [Figs. 5(a) and 5(b)] can be thermodynamically determined by the GFE density difference ($\Delta g = g_{C2/m} - g_{P2_1/c}$) under LPTL, which is a function of $I$ and $\omega_0$. The results are plotted as phase diagrams in Figs. 5(c) and 5(d). The red region indicates that the $P2_1/c$ is thermodynamically preferred, while the blue region gives a preference of $C2/m$. It can be seen that applying a 14 THz LPTL (regardless of polarization direction) would stabilize the $C2/m$ phase. In most cases the $P2_1/c$ is energetically favorable due to its being the ground state nature. In addition, the intensity of LPTL required to reverse the GFE of two phases is less than $10^9$ W/cm$^2$ (< 1 MV/cm), which can be experimentally achievable.[44,51] Due to the low absorption of THz light in the 2D system, a significant amount of light energy is wasted if light penetrates the thin film only once. Therefore, one may design a cavity shape structure or to select substrate that reflect the light in a confined space, which may increase the utilization of light energy.

To be specific, we set light intensity $I = 2\times10^9$ W/cm$^2$ (electric field peak of ~1 MV/cm) and plot the GFE densities of the two phases in Figs. 5(e) and 5(f). The corresponding saddle point transition structure results are also plotted, which could reflect the energy landscape under THz illumination. If the GFE of the TS lies in the middle of the two phases, barrier-free phase transformation (or at least a much reduced energy barrier) would be expected, which usually occurs in an ultrafast kinetics within a few picoseconds. For example, under 14 THz LPTL, the GFE satisfies $g_{C2/m} < g_{\text{TS}} < g_{P2_1/c}$, indicating a fast transition from $P2_1/c$ to $C2/m$. Under the $x$-LPTL at 4.5 THz (or $y$-LPTL at 10 THz), one has $g_{P2_1/c} < g_{\text{TS}} < g_{C2/m}$, so that the system would transform back to $P2_1/c$ quickly. In addition, the phonon lifetime value does not strongly affect the main conclusion (Figs. S9 and S10). This is because of that the phonon broadening has a significantly smaller magnitude (~0.1 THz) compared to the LPTL. The barrier-free transition happens wherever light is irradiated, and the conventional nucleation and growth process can be avoided.

In conclusion, we explore an ultrathin 2D RP-perovskite Bi$_2$FeO$_4$ with an all-optical read and write scheme. This material could exhibit two (meta-)stable phases, $P2_1/c$ and $C2/m$. Both of them are antiferromagnetic semiconductors, and are protected by sufficiently large energy barriers along phase transformation path. Owing to contrast band dispersions, they possess observable and distinguishable reflectance spectrum in the visible region. Through inclusion of both terahertz light absorption and scattering processes, we propose that they can be reversibly switched with a much-reduced energy barrier, realizing reversible and non-volatile phase transformations.

**Supplementary Material.** See the supplementary material for more details on the band structures, potential surface absorption by oxygen, and optomechanics theory under THz irradiation.

**Acknowledgments.** This work is supported by the National Natural Science Foundation of China under Grants Nos. 12374065, 11974270, and 21903063. The authors thank Chuanwei Fan for technical support on the computational resources and facilities. The high-performance computing (HPC) platform of Xi'an Jiaotong

University and Hefei Advanced Computing Center are also acknowledged.